\documentclass[review]{elsarticle}

\usepackage{hyperref}

\journal{Journal of \LaTeX\ Templates}

\begin{document}

\begin{frontmatter}

\title{Infrared and visible laser spectroscopy for highly-charged Ni-like ions}

\author{Yuri Ralchenko}
\address{Quantum Measurement Division, 
National Institute of Standards and Technology, Gaithersburg, MD
20899-8422, USA}
\ead{yuri.ralchenko@nist.gov}

\begin{abstract}

Application of visible or infrared (IR) lasers for spectroscopy of
highly-charged ions (HCI) has not been particularly extensive so far due to a
mismatch in typical energies. We show here that the energy difference between
the two lowest levels within the first excited configuration $3d^94s$ in Ni-like
ions of heavy elements from $Z_N$=60 to $Z_N$=92 is within the range of visible
or near-IR lasers. The wavelengths of these transitions are calculated within
the relativistic model potential formalism and compared with other theoretical
and limited experimental data. Detailed collisional-radiative simulations of
non-Maxwellian and thermal plasmas are performed showing that photopumping
between these levels using relatively moderate lasers is sufficient to provide a
two-order of magnitude increase of the pumped level population. This accordingly
results in a similar rise of the x-ray line intensity thereby allowing control of
x-ray emission with visible/IR lasers. 

\end{abstract}

\begin{keyword}
\texttt{Ni-like ions, visible and infrared lasers, spectroscopy}
\end{keyword}

\end{frontmatter}

Laser spectroscopy \cite{Demtroder} has become one of the most exciting and
successful fields of research. Its ubiquitous applications extend from
laser-induced breakdown spectroscopy (LIBS) to laser cooling of atoms and
molecules to optical lattices, to name a few. In practically all applications it
is the neutral or low-charge atoms (or molecules) that are the subject of
visible or infrared (IR) laser research. Since characteristic  energy
differences for ions rapidly grow with ion charge $z$, typically as $z$ (Coulomb
interactions) or even $z^4$ (spin-orbit interaction) for $\Delta n$=0
transitions (here $n$ is the principal quantum number) and $z^2$ for $\Delta n
\neq$0, common lasers with photon energies on the order of 0.5 eV to 3 eV cannot
be easily used for spectroscopy of mid- and high-charge ions. While the recently
developed x-ray free-electron lasers (XFEL) \cite{xfel} open a new chapter in
laser spectroscopy with highly-charged ions, the significant cost of XFELs
prohibits their widespread utilization in research.

Some highly-charged atomic ions may exhibit fine or hyperfine structure of the
ground configuration with energy differences within reach of traditional lasers.
One example is the B-like ion Ar$^{13+}$ with the ground configuration
$2s^22p$. Although the J=1/2 and J=3/2 levels of this configuration are
connected by a forbidden magnetic-dipole (M1) rather than an allowed
electric-dipole (E1) transition, a modest combination of a pulsed 100-Hz Nd:YAG
laser pumping a dye laser with a power of up to 1.2~W was sufficient to
photoexcite the 2.8-eV $2p_{3/2}$-$2p_{1/2}$ transition and accurately measure
its wavelength \cite{Mackel}. With the ion charge increase, as mentioned above,
this energy difference grows fast so that for B-like Fe$^{21+}$ it is about
14.7~eV, while for B-like Ba it reaches about 420~eV. Another example of
application of visible laser spectroscopy in highly-charged ions research is the
recent measurement of the hyperfine ground-configuration transitions in H-like
and Li-like ions of Bi \cite{Lochmann}.

In addition to the wavelength constraints, an effective interaction between
laser photons and an atom or an ion can only be reached when the lower
photoexcited state has a substantial population. Due to an infinite number of
converging atomic states (here we ignore continuum lowering effects in dense
plasmas), one can find transitions with almost any energy that is smaller than
the corresponding ionization potential. However, the populations of low excited
states are generally rather small due to possible E1 radiative decays with fast
rates that scale as $z^4$. The higher excited states in plasmas experience
strong collisions and therefore photoexcitation would be too weak to
affect their populations. The low metastable states certainly have larger
populations than other excited states, however, since forbidden radiative
probabilities increase with $z$ even faster than the E1 rates, in
highly-charged ions they may also be too depleted radiatively to be used in
laser photoexcitation studies. Fortunately, as will be discussed below, some
lowest metastable states in highly-charged ions have very small radiative decay
rates and thus their populations are extremely high.

\begin{figure}
\includegraphics[width=1.4\textwidth,angle=90]{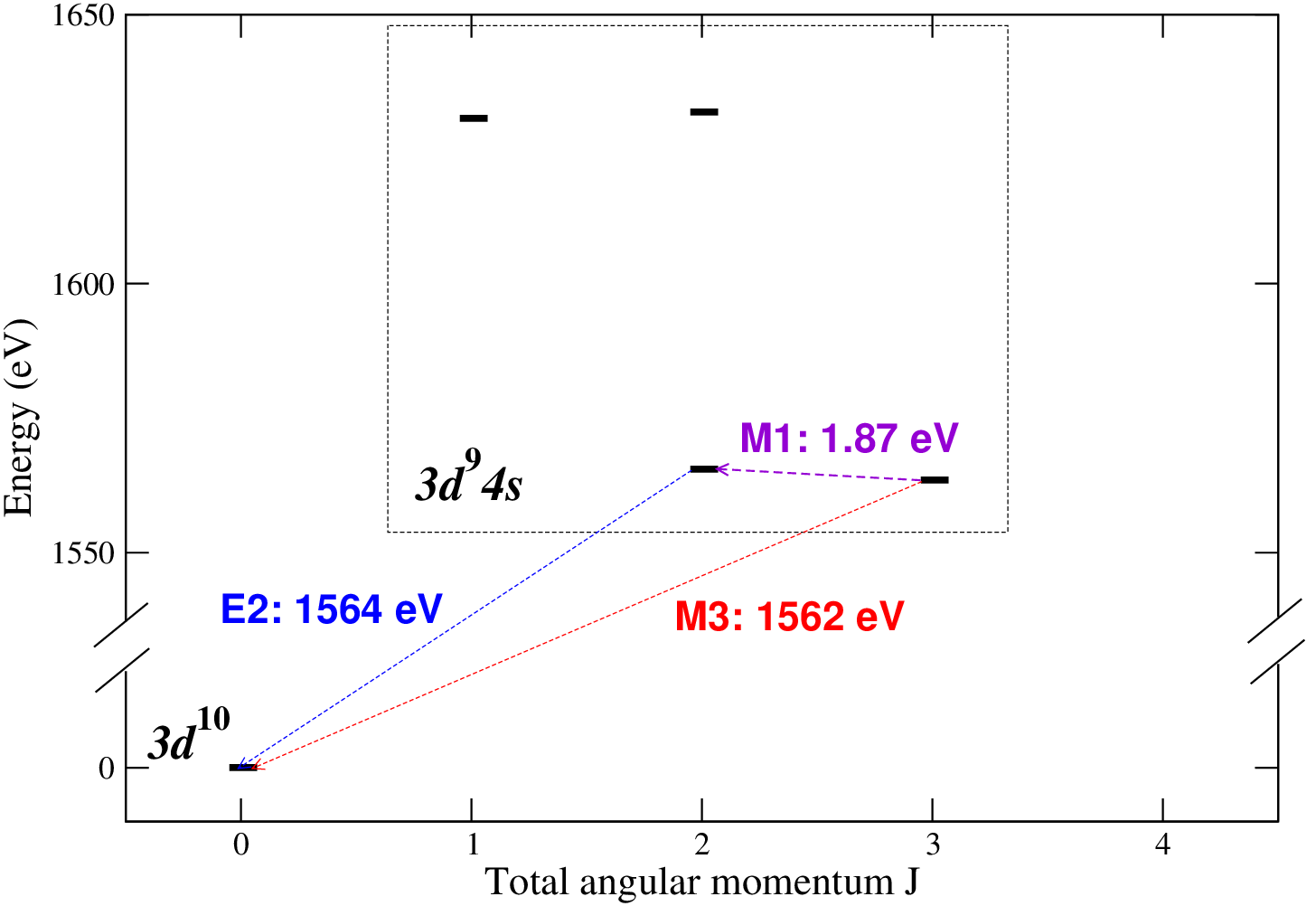}
\caption{\label{fig1}}
\end{figure}

In this paper we analyze applicability of visible and near-IR lasers to 
spectroscopy of highly-charged ions in the Ni-like isoelectronic sequence.
Although the ground configuration of neutral Ni is $3d^84s^2$, it becomes a
closed-shell configuration $3d^{10}$ for all other members of the isoelectronic
sequence \cite{ASD}. The first excited configuration above the single $J=0$
level of $3d^{10}$ is $3d^94s$ with the four levels having the following values
of the total angular momentum (in the increasing order of energies): 3, 2, 1,
and 2 (see Fig.~\ref{fig1} for the level structure of the $3d^94s$ configuration
in Ni-like W$^{46+}$). For highly-charged ions, this configuration clearly exhibits
$jj$-coupling, which results in doublet grouping of the 3-2 and 1-2 pairs of
levels. For instance, in Ni-like tungsten the energy difference between the two
doublets is about 40 times larger than their corresponding splittings
\cite{ASD}.

Since $3d^94s$ has the same parity as $3d^{10}$, there exist no allowed E1
transitions between these configurations. However, the forbidden $3l-4l'$
transitions of different multipoles become strong enough to overcome collisional
destruction in low-density plasmas and result in detectable spectral lines even
in laser-produced plasmas (see, e.g., \cite{Wyart}). Although
electric-quadrupole (E2) transitions have the highest decay rate, even the
magnetic-octupole (M3) line $3d^{10}$--$3d^94s$ (5/2,1/2)$_3$ was reliably
observed under low densities ($n_e \approx$ 10$^{11}$ cm$^{-3}$) of electron
beam ion traps (EBIT) already 25 years ago \cite{ThU_Ni}. Also, the M1 transitions from the J=1 $3d^94s$ levels have too small branching ratios to be of any importance for the following discussion.

Over the last decade the $3d^{10}$--$3d^94s$ E2 and M3 lines in Ni-like
W$^{46+}$ have been a subject of extensive analysis, in part due to importance
of tungsten for ITER research (e.g., \cite{ITER}). The low-resolution tokamak
and EBIT measurements \cite{put,PRA06} recorded an unresolved spectral feature
due to an overlap of these two close lines near 0.74~nm. The first attempt at
modeling W$^{46+}$ emission with the ADAS package \cite{put} was not
particularly successful in explaining the observed intensity although a more
detailed large-scale modeling of EBIT spectra \cite{PRA06} resulted in a much
better agreement with the measurements. It was explained later in
Ref.~\cite{Ral} that the J=3 and J=2 levels of $3d^94s$ that are connected by a
very weak M1 transition become strongly coupled via collisional-radiative
interaction with the intermediate configuration $3d^94p$. Consequently, the
population of the lower metastable J=3 level may be transferred to the J=2 level
via excitation followed by a radiative decay. Furthermore, the ratio of M3 and
E2 line intensities in Ni-like W was shown to be strongly dependent on electron
density in the range of values relevant to magnetic fusion devices \cite{Ral}.
The experimental work on these lines culminated with high-resolution
measurements at LLNL EBIT~\cite{W_Ni} that yielded observed wavelengths of 
0.79374~nm for M3 and 0.79280~nm for E2 with a better than 0.01\% accuracy. That
work also showed that the measured intensities of the M3 and E2 lines are
comparable in magnitude. Since radiative probabilities for these high-multipole
transitions differ by many orders (e.g., about $10^6$ for Ni-like tungsten
\cite{Ral}), it follows then that the  population of the J=3 level is larger
than that of the J=2 level by about the same factor. Therefore, one may expect
that photoexcitation of a very small fraction of population of the J=3 level
should result in a significant increase of the J=2 population and, accordingly,
the intensity of the E2 line. 

\begin{figure}
\includegraphics[width=1.4\textwidth,angle=90]{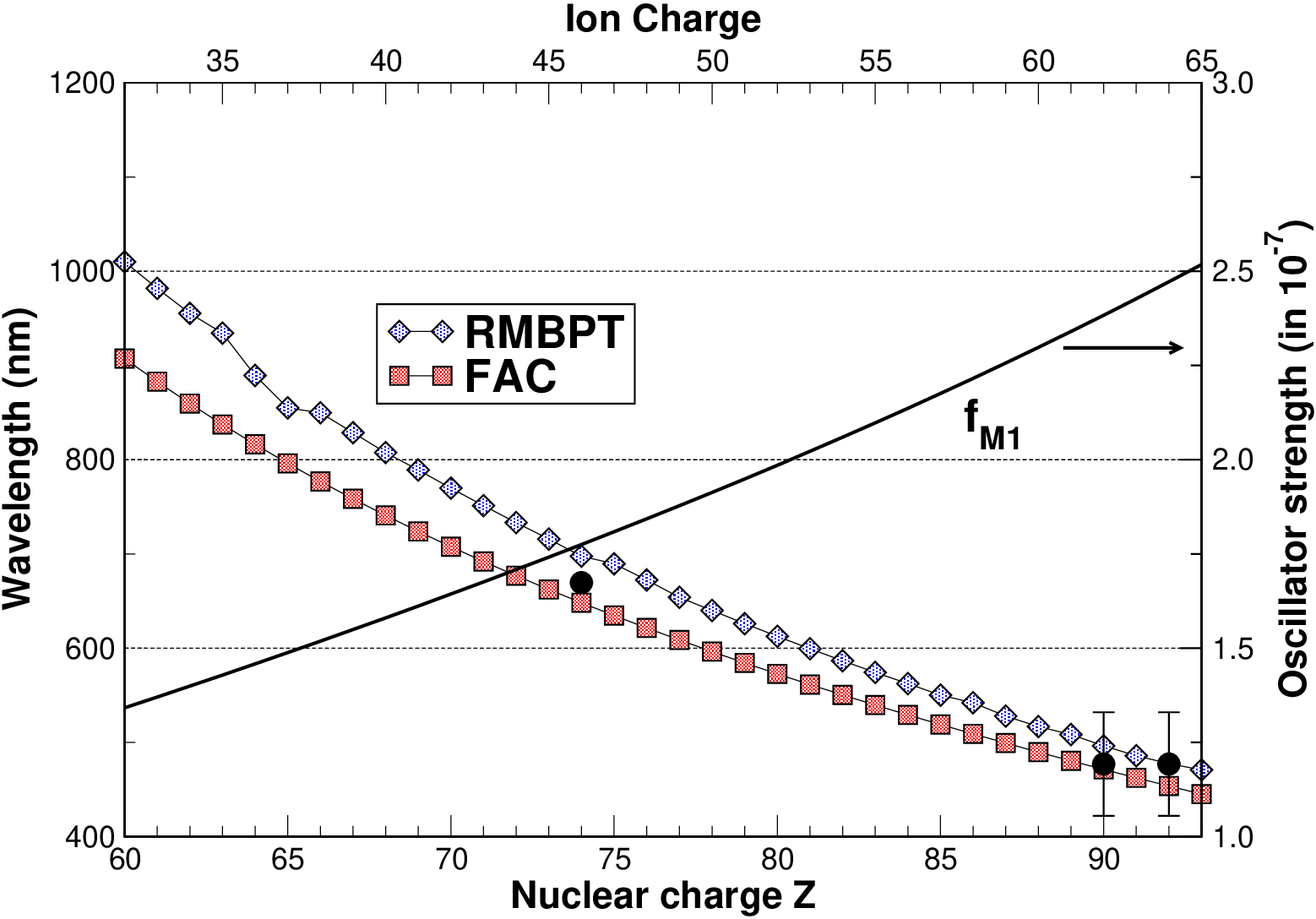}
\caption{\label{fig2}}
\end{figure}

While the energy differences for the J=3 and J=2 levels derived from the already
measured spectral lines in Ni-like W, Th, and U are in the visible range, it is
helpful to analyze their dependence along the isoelectronic sequence.
Figure~\ref{fig2} presents calculated and measured wavelengths for the J=3--J=2
M1 transition in Ni-like ions for the range of nuclear charges $Z_N$ = 60 to 93.
Our calculations were performed using the Flexible Atomic Code (FAC) \cite{FAC}
(squares). FAC implements the relativistic model potential method to calculate
wavefunctions and other atomic characteristics such as energies, radiative
transition probabilities, and electron-impact cross sections. Important
quantum-electrodynamic effects, e.g., Breit interaction, are added as well. The
relativistic many-body perturbation theory (RMBPT) calculations of
Ref.~\cite{Safronova} are presented by diamonds. The experimental data for W,
Th, and U are derived from the M3 and E2 wavelength measurements performed at
the LLNL EBIT \cite{ThU_Ni,W_Ni}. The uncertainties for the W data point are so
small that the error bars are within the symbol.

The measured and calculated wavelengths are seen to agree very well. Obviously,
the large uncertainties for the old Th and U data points make it more difficult
to quantify the disagreement but for W, the experimental and FAC wavelengths
agree within 3\%. This is a relatively large discrepancy for typical
atomic structure calculations but here the wavelengths are determined from the
differences of two large excitation energies, and thus the ensuing uncertainties
become larger. Nonetheless, it is clear that the wavelengths for this transition
are in the near-infrared and visible parts of spectra across the entire range of
nuclear charges from 60 to 93. The $z$ dependence of the wavelength is rather
weak, about $1/z$, which is typical for $\Delta n=0$ transitions.  The value of
the M1 oscillator strength shown in figure~\ref{fig2} (right y-axis) is rather
small, on the order of 2$\times$10$^{-7}$, which is not unexpected for forbidden
transitions with energy differences of a few eV. For comparison, the $f$-value
for the 441-nm Ar$^{13+}$ M1 photoexcitation transition \cite{Mackel} was only
about factor of 5 larger. Note also that the radiative probabilities for the
J=3--J=2 M1 transitions vary between 20~s$^{-1}$ and 120~s$^{-1}$, and thus they
are completely negligible in population kinetics for these levels.

\begin{figure}
\includegraphics[width=1.4\textwidth,angle=90]{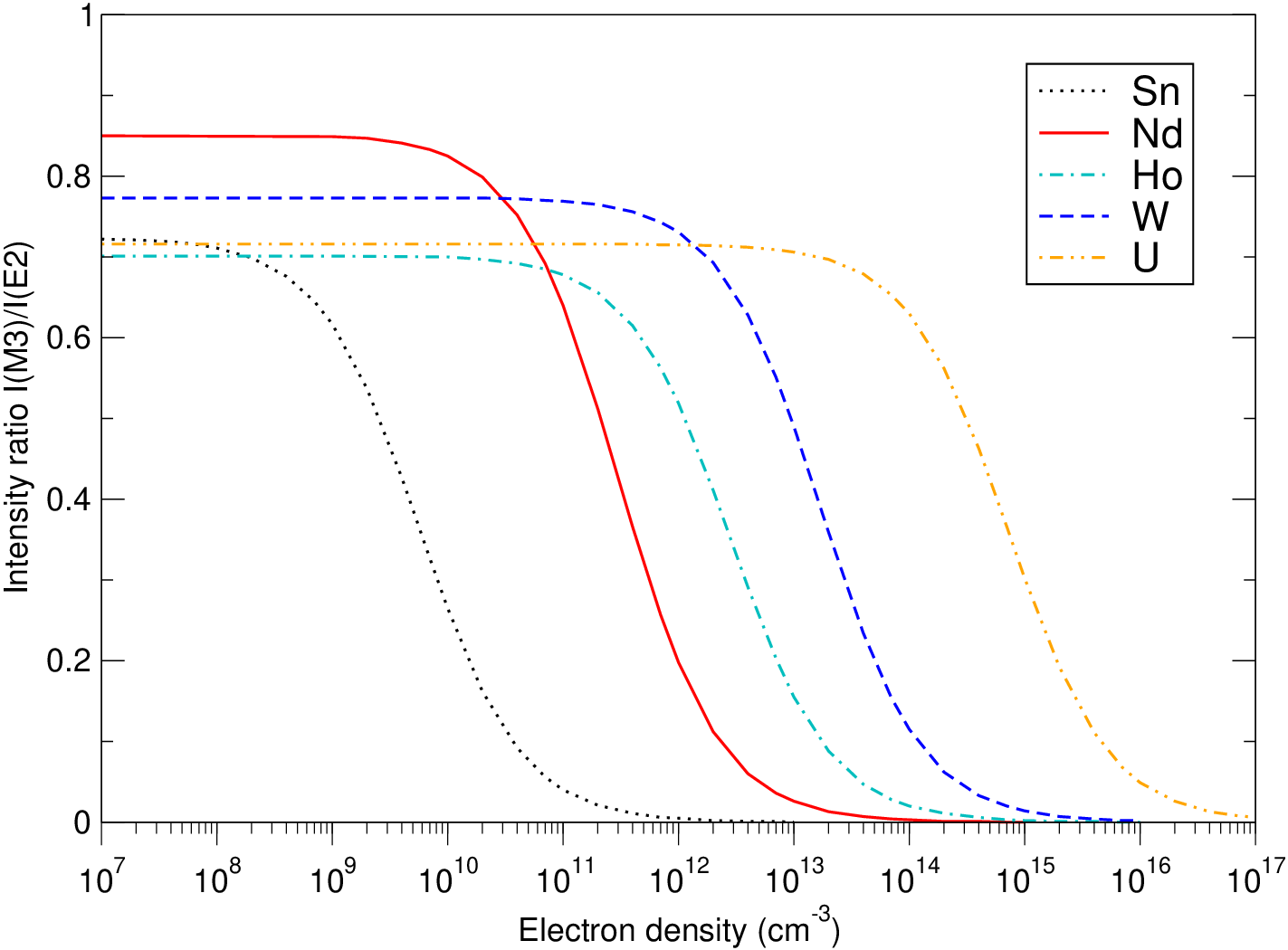}
\caption{\label{fig3}}
\end{figure}

While the wavelength of the M1 transition between the levels can be easily
matched by various lasers, the photoexcitation from J=3 to J=2 cannot be
effective above a specific value of electron density that varies along the
isoelectronic sequence. In Ref.~\cite{Ral} (see also \cite{ThU_Ni}) it was
shown in much detail that at higher densities a collisional-radiative (CR)
redistribution of population from J=3 into J=2 via $3d^94p$ levels efficiently
depletes the lower level. In figure~\ref{fig3} we present the calculated ratio
of the M3 and E2 line intensities for a number of Ni-like ions from $Z_N$=50 Sn
to $Z_N$=92 U in steady-state Maxwellian plasmas with temperatures of about the
ionization potential of the corresponding Ni-like ion. As discussed in
\cite{APiP}, for highly-charged high-Z elements the electron temperatures
required to reach maximal abundance for a specific ion are on the order or even
larger than its ionization potential. The CR simulations were performed with the
code NOMAD \cite{NOMAD,Book} using atomic data (energy levels, radiative
transition probabilities, electron-impact cross sections for excitation,
ionization, and recombination) calculated with FAC. The model includes Cu-, Ni-,
and Co-like ions which allows us to account for practically all important
physical processes affecting populations of the lowest excited levels in Ni-like
ions. The CR calculations show that at some value of the electron density (from
$10^8$~cm$^{-3}$ for Sn to $10^{14}$~cm$^{-3}$ for U) the electron-impact
excitation rate $3d^94s$--$3d^94p$ becomes comparable with the weak M3
transition probability and thus the J=3 level does not live long enough to reach
the corresponding radiative lifetime. In figure~\ref{fig3} this change in
primary depopulation process for J=3 corresponds to a sharp decrease of the
intensity ratio. Therefore, above these density limits the populations of the
lower level are generally too small to be used for photoexcitation.

In order to calculate the effect of laser photoexcitation on populations of the
J=3 and J=2 levels, one has to perform a full time-dependent CR modeling. Here
again the NOMAD code \cite{NOMAD} is applied to simulation for tungsten ions in
tokamaks and EBITs. The plasma parameters for EBIT conditions are
$n_e$=10$^{11}$~cm$^{-3}$ with an electron beam energy distribution represented
by a Gaussian function centered at 3950~eV and full width at half maximum (FWHM)
of 40 eV. For a Maxwellian tokamak plasma the corresponding parameters were
$n_e$=10$^{13}$~cm$^{-3}$ and electron temperature $T_e$=4200~eV. These values
of temperatures and beam energies provide significant abundance for Ni-like ions
of tungsten. Collisions with protons in tokamak plasmas may provide some
contribution to equilibration between the levels of interest that dissappears in
the low density limit; while this effect should in principle be clarified, here
we neglect it emphasizing the electron collisions. In both cases the simulation
starts from a steady-state condition at time $t$=0 and continues until
$t$=10$^{-2}$~s using a logarithmic grid of 150 steps. A laser with a wavelength
of 629.186~nm corresponding to the calculated energy difference, FWHM of 0.024
nm, and intensity of 10$^7$ W/cm$^2$ was  ssumed to illuminate the plasma from
$t$=10$^{-12}$~s either continuously until the final time or in a pulse mode
until 10$^{-8}$~s.

\begin{figure}
\includegraphics[width=1.4\textwidth,angle=90]{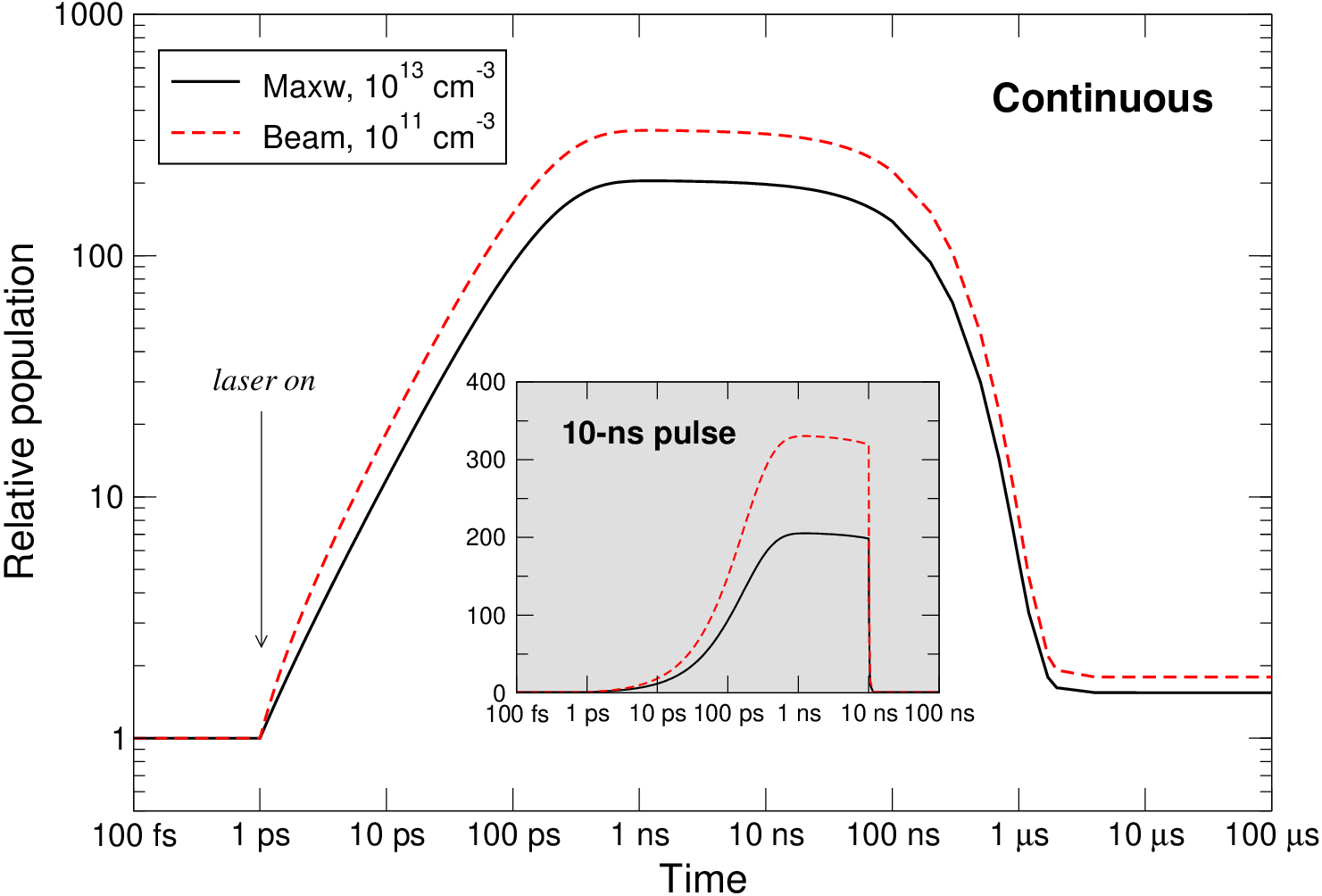}
\caption{\label{fig4}}
\end{figure}

The simulation results for the relative (normalized to the steady-state value at
$t$=0) population of the J=2 level are presented in figure~\ref{fig4}. As soon
as the laser is turned on, the photoexcitation J=3$\rightarrow$J=2 provides a
very strong population flux that increases the upper level population by more
than two orders of magnitude for both plasmas within just 1~ns. As mentioned
above, such a significant contribution is due to a very large imbalance of
populations between two levels. This equilibration time is on the order of the
radiative lifetime of the J=2 transition into the ground state which is the
fastest physical process among all possible depopulation mechanisms for this
level. The same lifetime gives an estimate of the relaxation time in the
pulse-mode simulation when the laser is turned off at 10 ns (inset in
figure~\ref{fig4}). The drop in population above 100~ns, in turn, is related to
depletion of the J=3 level due to photoexcitation with a rate of
4$\times$10$^6$~s$^{-1}$. At asymptotically large times, the relative population
limit reflects the established equilibration between the two levels when the
radiative decay of J=3 with the probability of about 1000~s$^{-1}$ is completely
negligible compared to laser photoexcitation. 

To summarize, the presented calculations show that the energy difference between
the lowest excited levels within the $3d^94s$ configurations in Ni-like high-Z
ions is within a few eV, which allows one to use it for photopumping with
visible or near-infrared lasers. Moreover, our time-dependent
collisional-radiative simulations for Ni-like W show that such photoexcitation
increases the population of the J=2 level by more than two orders of magnitude
and thus may result in observable detection of enhanced x-ray emission in the E2
transition into the ground state. Such a mechanism may allow control of x-ray
emission in hot plasmas of Ni-like ions with the standard laser techniques from
laser spectroscopy. Without loss of generality, we can assume that similar
effects can be observed in other Ni-like ions of heavy elements.

The author is grateful to U.I.~Safronova for providing detailed energy levels
for Ni-like ions from Ref.~\cite{Safronova} and to E. Stambulchik and A. Kramida
for valuable comments. The FAC input files used to calculate the atomic data are
available on request.

\newpage

List of figure captions:

Figure 1. (Colour online.) Levels of the $3d^94s$ configuration in Ni-like
W$^{46+}$ \cite{ASD}. The doublet structure typical for $jj$-coupling is clearly
visible. The transition energy of 1.87~eV between the J=3 and J=2 levels
corresponds to the photon wavelength of 663~nm.

Figure 2. (Colour online.) The calculated (present work as squares and
\cite{Safronova} as diamonds) and measured (solid circles) \cite{ThU_Ni,W_Ni}
wavelengths for the J=3--J=2 transitions in Ni-like ions. The corresponding
magnetic-dipole oscillator strengths (solid line, right y-axis) are shown in
units of $10^{-7}$.

Figure 3. (Colour online.) The intensity ratios between M3 and E2 transitions in
Ni-like ions of Sn, Nd, Ho, W, and U as functions of electron density.

Figure 4. (Colour online.) Relative populations of the J=2 level (with regard to
the steady-state population) under continuous laser photopumping starting at
t=10$^{-12}$~s. Solid line: Maxwellian plasma at 4200~eV and electron density
$n_e$ = 10$^{13}$ cm$^{-3}$; dashed line: Gaussian electron energy distribution
at 3950~eV and $n_e$ = 10$^{11}$ cm$^{-3}$. Inset: same plasma conditions but
the laser is turned off at 10$^{-8}$ s.

\newpage

\newpage

\newpage

\newpage

\end{document}